\documentstyle[amstex,epsf,psfig]{mn}
\newcommand{\I}[1]{{\mathcal I } }
\newcommand{\bm}[1]{\boldmath{#1}}

\title{The kinematic Sunyaev Zeldovich effect and 
transverse cluster velocities}

\author[E. Audit and J.F.L. Simmons]
{Edouard Audit$^{1}$, John F.L. Simmons$^{2}$ \\
  $^{1}$ Laboratoire d'Astrophysique Extragalactique et de Cosmologie
   -  CNRS URA 173 - \\
  Observatoire de Meudon - 5, Place Jules Jansen - 92
  195 Meudon - France\\
  $^{2}$ Department of Physics and Astronomy, University of Glasgow
  Glasgow G12 8QQ - U.K}

\date{Accepted 199-
      Received 199-
      in original form 199-}

\pubyear{1997}

\begin{document}
\onecolumn

\maketitle

\begin{abstract}
The polarization  of the CMBR scattered  by galaxy  clusters in the kinematic
Sunyaev Zeldovich effect depends on  the transverse velocity of the  cluster.
This polarizing effect is  proportional  to the transverse  velocity squared,
and so  weaker that the change  in intensity due  to the radial motion in the
kinematic  effect.  The value given by  Sunyaev  and  Zeldovich, and which is
frequently cited,  underestimates the polarizing effect  by  a factor of ten.
We   show furthermore  that   the   polarization  has   a   strong  frequency
dependence.  This means that the  polarization should be  detectable with the
new generation of CMBR probes,  at least for some  clusters. Thus this effect
offers,  almost uniquely, a method  of  obtaining  the vectorial velocity  of
clusters.
\end{abstract}

\begin{keywords}
  Cosmology: theory -- cosmic microwave -- polarization
\end{keywords}

\section{Introduction}

One of the  main goals of observational cosmology  in the coming years is the
measurement of the CMBR and its fluctuations. Before arriving to us, the CMBR
photons traverse the  entire observable universe. It  follows that if one can
observe the effect of the interaction of these photons  with matter they pass
through,  then  one should be able   to obtain crucial information  about the
distribution of matter in  the universe. With  this perspective,  Sunyaev and
Zeldovich studied the  various scattering processes  that  take place between
black body  photons and electrons in galaxy  clusters. These phenomena, which
are referred to collectively as the 'Sunyaev-Zeldovich effect', are now being
studied  in  great detail, and indeed  are  being used to  probe cosmological
structure formation (\cite{birk} and reference therein).

In their  paper    of 1980  Sunyaev and Zeldovich    \cite{SZ}  discussed the
possibility of  measuring the radial velocity of  a cluster of  galaxies from
the enhancement of the CMBR temperature in the direction of the cluster. This
effect is somewhat   similar to the   well known Sunyaev Zeldovich  effect in
which CMBR photons  are Compton scattered  by the hot  gas in  the cluster, a
process which gives  rise  to  a distortion  in  the  CMBR spectrum in    the
direction of the cluster.  For this reason the  effect due to the bulk motion
of the cluster has become known as the kinematic S-Z effect, and the original
effect as the thermal S-Z effect. The thermal effect leads to a variation
in intensity of the radiation given by

\begin{equation}
\label{sz2}
\frac{\Delta I_\nu} {I_\nu}= 2y \frac{x\exp {x}}{\exp {x}-1 }\;\frac{x}{2\coth
x/2}-2  \ ,
\end{equation}
where  $y=\tau  kT_e  /m_e  c^2$  is the    comptonization factor, $T_e$  the
temperature of the  electrons in the cluster,  $x=h\nu /kT_o$,  $\tau$ the
optical depth, and $T_o$ the temperature of the CMBR. For small values of $x$
the change in intensity is negative, whilst for large values it is positive.

The kinematic effect gives rise to a change in intensity given by
\begin{equation}\label{sz4}
\frac{\Delta I_\nu} {I_\nu}= -\beta_r \tau \frac{x\exp {x}}{\exp {x}-1 }
\ ,
\end{equation}
where $\beta_r$ is the radial  velocity in units of  the speed of light.  The
sign  of   the fractional change   in intensity  does  not  change  sign with
frequency, and  is  positive if the  peculiar  radial motion  is  towards the
observer. The  effective temperature change  in independent of frequency, and
given by $\Delta T/T = -\beta_r \tau.$

In the same paper the question of the polarization  that can be produced when
the motion of the cluster has a component transverse to  the line of sight is
also  discussed, and the degree of   polarization induced in integrated light
from the cluster is claimed to be  of the order $ 0.1  \beta_t ^2 \tau$ where
$\beta_t$ is the transverse velocity of the cluster  relative to the CMBR. In
fact, as  we show below, the  effect is exactly ten  times larger,  and has a
strong frequency  dependence.  (Sunyaev and  Zeldovich also argued that there
would be a contribution from  higher order scattering  given by $1/40 \beta_t
\tau^2$, which  could be of a comparable  size for larger optical depths, but
we shall not discuss this here as the  latter effect is still more negligible
than  thought  previously.)  This   polarization  will   be in   a  direction
perpendicular to the plane formed  by the line  of sight and the velocity  of
the cluster.  As Sunyaev and Zeldovich point out, potentially this polarizing
effect of the kinematic  S-Z  effect can yield  a  great deal of  information
about  the  peculiar velocities  of galaxy  clusters,   and indeed in  theory
represents  one  of the few ways   of  determining their  vectorial velocity.
Peculiar   velocities  are thought   to   lie  in  the   range  500-3000   km
s$^{-1}$. Cluster optical depths are   badly known but  appear  to be in  the
range  0.01 to 0.05, yielding  a frequency integrated degree polarization due
to single scattering of the  order $2\ 10^{-7}$ to  $5\ 10^{-6}$.  Such small
values of polarization could be still be difficult to detect.

However,  the   frequency dependence  of  this   polarization  is of  crucial
importance, firstly because this  frequency dependence helps  distinguish the
signal   from  other   polarizing    mechanisms,  and  secondly   because the
polarization rapidly increases as a  function of frequency. Although  Sunyaev
and Zeldovich  did point out that  the  polarization should  have a frequency
dependence,  they calculated only the frequency  integrated polarization.  In
this  paper we look  more closely at   the dependence of  the polarization on
frequency and find that at values of $x>5$ the polarization will be orders of
magnitude higher than the frequency integrated polarization. Furthermore this
mechanism has  a characteristic frequency dependence  which should enable one
to distinguish it from other polarizing mechanisms.

\begin{figure}
\psfig{file=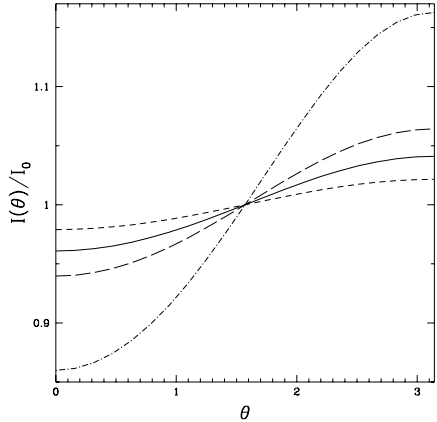,height=6.cm}
\caption{The intensity  of  the CMBR  as   seen by the  cluster  at different
frequencies plotted against polar viewing angle, $\theta$. The
velocity of the cluster, $\beta$, is taken to be 0.01. The
small-dashed curve corresponds to 100 GHz, the long-dashed to 353
GHz and the dotted to 857 GHz (these frequencies correspond to
three of the Planck-HFI instrument). The solid curve corresponds to
the total (frequency integrated) intensity.}
\label{anisotropy}
\end{figure}

The origin of the polarization in the case of the kinematic S-Z effect is the
apparent  anisotropy of  the  CMBR radiation seen by  the  cluster due to its
motion relative  to  the CMBR frame.   Although the  degree  of anisotropy in
frequency integrated light for  low values of $\beta$  is not very large, the
monochromatic  intensity   displays   much   greater   anisotropy.     Figure
[\ref{anisotropy}] shows the variation of incident intensity as a function of
the viewing  direction, described by the  polar angle, $\theta$, with respect
to  the axis  of symmetry  of the  radiation   field (defined by the  cluster
velocity), at different frequencies and for $\beta  = 0.01$. Now consider the
case   where    the  cluster is   moving   perpendicularly   to  the  line of
sight. Photons incident from the forward direction  and scattered towards the
observer   will  be scattered  through   $90^o$, and  hence  fully  polarized
perpendicular to the  line of sight and velocity   of the cluster.  Similarly
photons  from the  backward    direction  will  be  polarized   in  the  same
direction. Photons arriving perpendicular to  this plane will be polarized in
the opposite sense, and  hence tend to  cancel out the  effect of the photons
scattered from the forward and  backward direction. However the intensity  of
the incident radiation varies with the angle of incidence, $\theta$. It is in
fact the quadrupole that produces the net polarization.

We  also  find that  the   degree of  polarization depends  not   only on the
transverse velocity, but also weakly on  the absolute velocity of the cluster
with respect to the CMBR in a non-degenerate way. In principle this means that
measurement of the degree of polarization at different frequencies provides a
means of  determining the vectorial velocity  without reference either to the
kinematic  intensity  SZ effect, and independently   of any knowledge  of the
optical depth. In practice, however, use of  optical depth estimates from the
thermal SZ effect, and the radial component of  the cluster velocity from the
kinematic intensity  SZ effect  would provide much  greater precision  on the
vector velocity measurement.

As in the  case of the Sunyaev  Zeldovich paper, we  shall throughout use the
Thomson scattering approximation, using  relativistic corrections for Doppler
and aberration effects.  Of  course, a complete  treatment should  consider a
distribution of hot electrons, in which  case the relativistic expression for
electron scattering should be used. For the points  we make in this paper the
simplified treatment  is justified. Moreover,  by formulating  the problem in
the cluster  rest  frame and including  all relativistic  transformations and
directional effects, this treatment  makes itself amenable to generalisation
to the full relativistic case.

\section{Analytic expressions for the Stokes parameters  }

\begin{figure}
\hspace*{1.0cm}\psfig{file=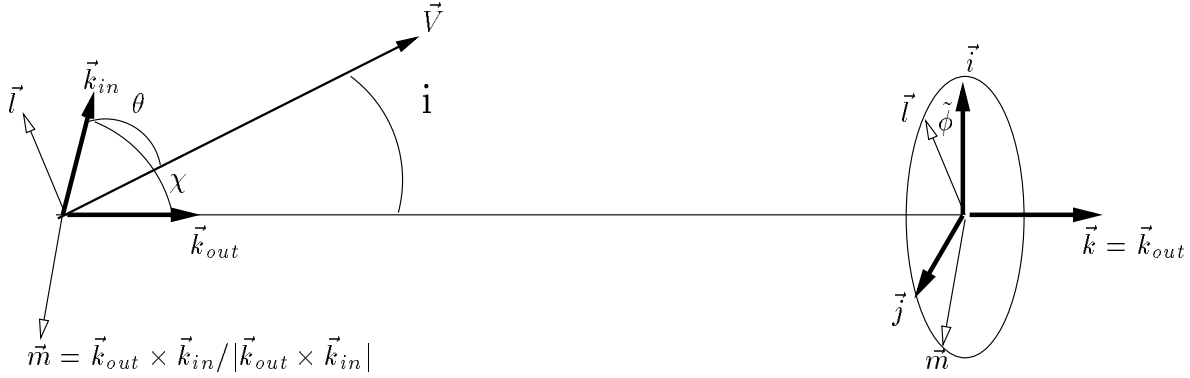,height=5.cm}
\caption{Scattering    geometry:  Unpolarized light  incident  on  cluster is
scattered (by  stationary  electrons) towards  the observer through  an angle
$\chi$. The scattering plane is defined by the  cluster velocity $\vec V$ and
the line of sight.}
\label{geo}
\end{figure}

The  geometric situation is represented in  Figure  [\ref{geo}].  The cluster
has velocity  $\vec{V}$ relative to  the rest  frame of  the CMBR, ${\mathcal
F}'$. We shall assume throughout that the CMBR  has a perfect isotropic black
body spectrum. The cluster will see an anisotropic but axisymmetric intensity
distribution.  Let $O$ be an observer at our position but at rest in the
frame of the cluster, ${\mathcal F}$, and $i$ be the angle between the axis of
symmetry and the line of sight.  We will first  compute the Stokes parameters
as seen by $O$.  (All the primed quantities  correspond the the rest frame of
the CMBR and the unprimed to the rest frame of the cluster).

The intensity of the CMBR in ${\mathcal F}'$ has  a Planck distribution given
by:
\begin{equation}
{\mathcal I}'(\nu') = A \frac{\nu^{\prime 3}}{e^{h\nu'/kT_0} - 1}
\ .
\end{equation}

Consider  an incident   photon in  direction $\vec  k_{in}$   (with polar and
azimuthal angle $\theta$ and $\phi$ in the  cluster frame ${\mathcal F}$) and
frequency $\nu$ also in  the cluster frame, where  the $z$ axis is defined by
the axis of symmetry (defined by $\vec{V}$).  The frequency of this photon in
${\mathcal F}'$ will be given by $\nu' = \nu  (1 + \beta \cos{\theta})\gamma$
where $\gamma = (1 -  \beta^2)^{-1/2}$.   Due to relativistic abberation,  the
apparent direction of   the photon will  be  different in ${\mathcal  F}$ and
${\mathcal F}'$, but it does not matter since the radiation field is isotropic
in ${\mathcal F}'$.   Writing the intensity  in ${\mathcal  F}$ as ${\mathcal
I}(\nu  , \theta)$  and in ${\mathcal  F}'$ as  ${\mathcal I}'  (\nu')$,  and using the
invariance property of the intensity, $ {\mathcal I}  / \nu^3 = {\mathcal I}'
/ \nu^{\prime 3}$ we have
\begin{equation}
{\mathcal I}( \nu, \theta)
= \left(\frac{\nu}{\nu'}\right)^3 {\mathcal I}'(\nu')
= A \frac{\nu^{3}}{e^{h\nu /kT(\theta)} - 1} \  ,
\end{equation}
where  $T(\theta) (1  +   \beta  \cos{\theta})\gamma =  T_0$.  The  frequency
integrated intensity in the cluster frame in a direction $(\theta, \phi)$ is
given by
\begin{equation}
{\mathcal I}(\theta) \propto T^4
= \frac{T_o^4}{(1+\beta\cos{\theta})^4\gamma^4} = (T_o  / \gamma)^4 \left( 1 -
4\beta\cos{\theta} + 10 \beta^2 \cos^2{\theta} + {\mathcal O}(\beta^3)\right)
\ .
\label{temp}
\end{equation}

The  incident and  scattered  photons will  have  the same  frequency  in the
cluster frame.     The scattering angle   is  given  by  $\cos \chi   = {\vec
k}_{in}\cdot {\vec  k}_{out}  $, and  the  unit vector perpendicular   to the
scattering plane   by  ${\vec m}=({\vec  k}_{out}\times  {\vec k}_{in})/|{\vec
k}_{out}\times {\vec   k}_{in}| $.   The  right handed   orthonormal  basis is
completed by ${\vec l}$.

The polarization  is best described by  the Stokes parameters $({\mathcal I},
Q, U, V)$  which represent the  total intensity, the difference  in intensity
measured  by  polarimeter  in two   given perpendicular  directions, the same
difference but with the polarimeter rotated through  $45^o$, and the circular
polarization  respectively  \cite{chandra}.   The    incident beam   will  be
unpolarized. Since Thomson scattering produces  no circular polarization, $V$
can be ignored.

We take an observer  basis ${\vec i},{\vec  j},{\vec k}$, where ${\vec k}$ is
equal to the direction $\vec k_{out}$ of the scattered photon, and ${\vec j}$
is perpendicular to  the plane defined   by $\vec{V}$ and  the  line of sight
(${\vec k}$).  This defines the orientation  of the polarimeter. To calculate
the contribution   of all  incident photons   we  have  to  rotate   from the
scattering plane to the observer frame, and  integrate over all solid angles,
$\Omega$, and along the line  of sight. We denote  the angle of this rotation
by  $\tilde{\phi}$ (see   figure [\ref{geo}]).   Using  the phase  function for
Thomson  scattering, and introducing  for convenience the abbreviations $S_x$
(resp. $C_x$) to denote the sine (resp.  cosine) of the  variable $x$, we may
write
\begin{equation}
{\mathcal I}_{sc}(\nu) = \frac{3 \tau}{16 \pi}
\int {\mathcal I} (\nu, \theta) (1 + C^2_\chi) C_{2 \tilde{\phi}} \;
d\Omega \  , \;\;\;\;\;\;\;
Q(\nu) = \frac{3 \tau}{16 \pi}
\int {\mathcal I}(\nu, \theta)  S_\chi^2 C_{2 \tilde{\phi}} \;
d\Omega \ , \; \; \;
U(\nu) = \frac{3 \tau}{16 \pi}
\int {\mathcal I}(\nu, \theta)  S^2_\chi S_{2 \tilde{\phi}} \;
d\Omega  .
\label{eq1}
\end{equation}
In  order to obtain the  frequency  integrated Stokes parameters,  ${\mathcal
I}(\nu, \theta)$ should  be replaced by  ${\mathcal I}(\theta)$ in  equations
(\ref{eq1}). The optical thickness, $\tau$, is given by an integral along the
line  of sight $\tau  = \int  \sigma_T n_e  dl$, where $n_e$  is the electron
density  and $\sigma_T$ the  Thomson cross-section.    Noting that $C_\chi  =
C_\theta C_i - S_\theta C_\phi  S_i $, $C_{\tilde{\phi}}  = -( C_\theta S_i +
S_\theta C_\phi  C_i)/S_\chi$ and  $S_{\tilde{\phi}}    = S_\theta S_\phi   /
S_\chi$ and integrating over $\phi$, equations (\ref{eq1}) reduce to
\begin{equation}
{\mathcal I}_{sc}(\nu) = \frac{3 \tau }{8}
\int_{-1}^{1}  {\mathcal I}(\nu,  \theta)
(1+\mu^2 - S_i^2 (3\mu^2 -1)/2) d\mu ,
\;\;\;\;\;\;\;
Q(\nu) = \frac{3 \tau}{16}  S_i^2
\int_{-1}^1   {\mathcal I}(\nu,  \theta)  (3\mu^2 -1)   \; d\mu ,
\;\;\;\;\;\;\;
U(\nu) = 0 .
\label{eq2}
\end{equation}
The total intensity seen by $\bm{O}$ is then given by
\begin{equation}
{\mathcal  I}_{tot}(\nu)
= {\mathcal I} + {\mathcal I}_{sc} - {\mathcal I}_{ab} \ ,
\end{equation}
where ${\mathcal  I}={\mathcal I}(\nu,  i)$  is the direct light,  ${\mathcal
I}_{sc}$ is the light scattered toward the observer  and ${\mathcal I}_{ab} =
\tau {\mathcal I}$  is the light scattered out   of the line  of  sight.  The
degree of polarization, $p$, of the net light seen  by $\bm{O}$ is then given
by
\begin{equation}
p(\nu) = |Q(\nu)|/ {\mathcal I}_{tot}(\nu) \ .
\end{equation}

All  the  above quantities  are expressed for  the observer  $\bm{O}$. For an
observer $\bm{O}'$ at rest in the CMBR frame we  have $p'(\nu') = p(\nu)$ and
${\mathcal  I}'(\nu')  = (\nu'/\nu)^3 {\mathcal I}(\nu)$   where  $\nu' = \nu
(1+\beta C_i)\gamma$.

The  fraction change  in intensity  given  by  $\Delta   {\mathcal I}(\nu)  =
({\mathcal I}_{tot}(\nu)  -  {\mathcal I}(\nu)) /{\mathcal  I}(\nu)$ describe
the kinematic SZ  effect and, of course, coincide  with the result of Sunyaev
and Zeldovich \cite{SZ} in the case of radial motion $S_i = 0$.

We can now compute the frequency  integrated degree of polarization by simply
replacing ${\mathcal I}(\nu, \theta)$  by ${\mathcal I}(\theta)$ in equations
(\ref{eq2}).   Keeping  only  terms up   to  $\beta^2$  in  the  expansion of
${\mathcal I}(\theta)$ (eq.   \ref{temp}) and  taking
${\mathcal  I}_{tot}  = {\mathcal I}$ and we obtain

\begin{equation}
p = S_i^2 \beta^2 \tau = \beta_t^2 \tau \ ,
\end{equation}
where we  have introduced the  transverse velocity $\beta_t$. This results has
the same velocity and optical depth dependence as the one obtained by Sunyaev
and Zeldovich, but is ten time greater.   Since $p$ is Lorentz invariant this
old also in  the  CMBR frame.   Furthermore, as  we  shall  see in   the next
section,  the  polarization  has a   strong frequency dependence.     In the low
frequency limit, the  degree of polarization is  independent of the frequency
and is given by

\begin{equation}
p = S_i^2\frac{3}{16}\tau \left(
\frac{(3-\beta^2)}{\beta^3}\ln\left(\frac{1+\beta}{1-\beta}\right)
-\frac{6}{\beta^2} \right) \simeq 0.1 \beta_t^2 \tau  + {\mathcal O}(\beta^4)
\ .
\label{lim_low}
\end{equation}
This results is similar to what was claimed  by Sunyaev and Zeldovich for the
frequency integrated effect.

\section{Numerical results }

\begin{figure}
\psfig{file=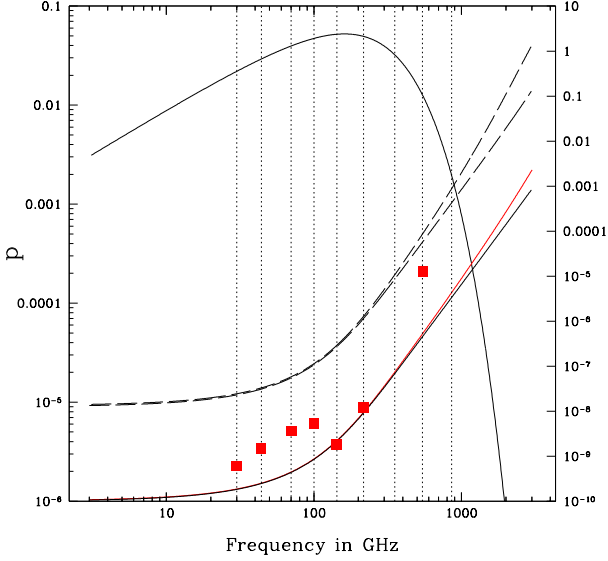,height=8.cm}
\caption{The degree of polarization is  plotted against frequency in GHz. The
transverse velocity of the  cluster is $\beta_t = 0.01$ for the lower  group
(solid line) of curves and $\beta_t  = 0.03$ in the upper  (dashed line).  In
each group the lower curve corresponds to $i=90^o$ and the upper to $i=45^o$.
The solid upper curve  shows  the intensity of   the CMBR in arbitrary  units
(right vertical axis).     The vertical dotted    lines shows the   different
frequency bands of the Planck experiment and the squares display the expected
sensitivity of the individual polarimetric detectors (when they exist). }
\label{pol_nu}
\end{figure}

In this section we present the degree of polarization that should be observed
for a typical galaxy cluster. All our results are  given for $\tau = 0.1$ but
values for other optical depths can simply be scaled linearly.

Figure [\ref{pol_nu}] displays  the degree of polarization  against frequency
for various transverse cluster speeds.  We can see  that $p$ slightly depends
on  the inclination even  for  constant values   of $\beta_t$.   As mentioned
previously, the  polarization reaches a  constant value in  the Raleigh-Jeans
(low frequency) limits  according to formula [\ref{lim_low}].   The variation
of $p$ with frequency is very strong. At $1000  GHz$ the polarization is more
than  two orders of  magnitude greater than the  lower  limit.  In the Planck
higher frequency band, the  expected polarization for $\beta=0.01$ and $0.03$
are of $0.01 \%$ and $0.10\%$  respectively, although these need to be
scaled according   to the optical depth.   Such   degrees of polarization are
above  the    detection limits   foreseen by    coming   experiments.  Figure
[\ref{pol_nu}] display    the  sensitivity  of  the   individual polarimetric
detector of the Planck experiment. Since there are between 8 and 34
polarimetric detectors per band and since a cluster could have an angular size
sensibly larger than the instrument resolution, it  should be possible to get
the  polarimetric signal by proper   processing of the  data. Furthermore, new
experiments could be designed especially for polarimetric measurement.

\begin{figure}
\psfig{file=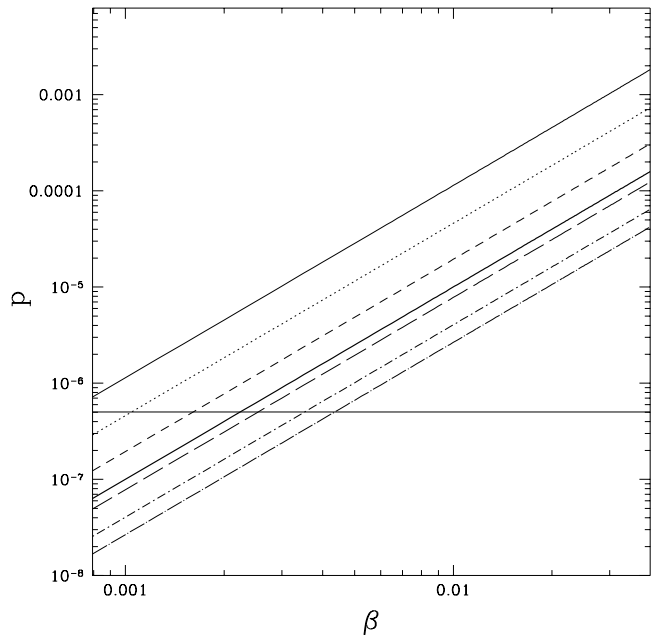,height=9.cm}
\caption{The degree of polarization   is plotted against $\beta$  for  purely
transverse motion ($i=90^o$). Each curve corresponds to one of the Planck-HFI
band: 857 GHz (solid  line),  545 GHz (dotted line),   353 GHz (short  dashed
line), 217 GHz (long dashed line), 143 GHz (dot - short dashed line), 100 GHz
( dot - long dashed line).  The  solid horizontal lines represents the degree
of primordial polarization expected for  a standard CDM   model (Hu \&  White
1997) and  the   thick  solid  line the   frequency   integrated   degree  of
polarization.}
\label{pol_beta}
\end{figure}

Figure [\ref{pol_beta}] displays the  degree of polarization  against $\beta$
for different frequencies. For all frequencies, as for the integreted effect,
the degree of polarization has a  $\beta^2$ dependence.  We  can see that the
polarization due to Thomson scattering in the cluster is much larger that the
expected  primordial polarization for  a wide  range  of parameter and should
therefore be measurable.


\section{Conclusions }
Transverse velocities of galaxy clusters will produce polarized signal in the
scattered CMBR.   This  polarization will  be  perpendicular  to the  cluster
velocity and the light of sight.  Although in  frequency integrated light the
level  of polarization   might be difficult   to detect  with the  techniques
presently available, the   polarization  has a strong  frequency  dependence,
increasing  towards higher frequencies. For  values of transverse velocity of
1000 km s$^{-1}$, and an optical depth of 0.02, the degree of polarization at
857 GHz is around  $2 . 10^{-6}$.  Since  the polarization depends on the
square of the transverse velocity, and linearly on the optical depth, certain
clusters should show  higher polarization. Other  sources of polarization, in
particular dust   and     synchrotron  radiation,   will   contaminate    the
signal.   However  filtering   techniques   that   take   into account    the
characteristic frequency dependence of  this polarization should allow one to
extract the signal due to the kinematic  effect.  Measured polarization could
then be  used to infer transverse  velocities of clusters. This question will
be elaborated in future work.

\end{document}